\documentclass[sigconf]{acmart}
\renewcommand\footnotetextcopyrightpermission[1]{}
\usepackage{amsmath}
\usepackage{amsfonts}
\usepackage{graphicx}
\usepackage{adjustbox}
\usepackage{multirow}
\usepackage{algpseudocode}
\usepackage{bm}
\usepackage{url}
\setcitestyle{numbers,sort&compress}
\newcommand{\mathbi}[1]{\textbf{\em #1}}
\newcommand{\R}{\mathbb{R}}

\acmConference[CIKM' 17]{International Conference on Information and Knowledge Management}{November 6--10}{Singapore}

\title{Spectrum-based deep neural networks for fraud detection}

\author{Shuhan Yuan}
\affiliation{Tongji University}
\email{4e66@tongji.edu.cn}
\author{Xintao Wu}
\affiliation{University of Arkansas}
\email{xintaowu@uark.edu}
\author{Jun Li}
\affiliation{University of Oregon}
\email{lijun@cs.uoregon.edu}
\author{Aidong Lu}
\affiliation{University of North Carolina at Charlotte}
\email{aidong.lu@uncc.edu}

\begin{document}

\begin{abstract}
In this paper, we focus on fraud detection on a signed graph with only a small set of labeled training data. We propose a novel framework that combines deep neural networks and spectral graph analysis. In particular, we use the node projection (called as spectral coordinate) in the low dimensional spectral space of the graph's adjacency matrix as input of deep neural networks. Spectral coordinates in the spectral space capture the most useful topology information of the network. Due to the small dimension of spectral coordinates (compared with the dimension of the adjacency matrix derived from a graph), training deep neural networks becomes feasible.
We develop and evaluate two neural networks, deep autoencoder and convolutional neural network, in our fraud detection framework. Experimental results on a real signed graph show that our spectrum based deep neural networks are effective in fraud detection.
\end{abstract}


\maketitle

\section{Introduction}

Online social networks (OSNs) have become popular social services for linking people together. Unfortunately, due to the openness of OSNs, fraudsters can also easily register themselves, inject fake contents, or take fraudulent activities, imposing severe security threads to OSNs and their legitimate participants.
Many fraud detection techniques have been developed in recent years \cite{Akoglu2012What,Jiang2014Catchsync,Cao2014Uncovering,Ying2011Spectrum}, including content-based approaches and graph-based approaches.
Different from content-based approaches that extract content features, (i.e., text, URL), from user activities on social networks \cite{Benevenuto2010Detecting}, graph-based approaches identify frauds based on network topologies. Often based on unsupervised learning, the graph-based approaches consider fraud as anomalies and extract various graph features associated with nodes, edges, ego-net, or communities from the graph \cite{Noble2003GraphBased,Akoglu2015Graph}.

In practice, a small set of labeled users are often available and hence supervised learning based detection approaches could be developed.
In this paper, we introduce deep neural network models for detecting frauds in signed graphs. Deep neural networks have achieved remarkable results in computer vision, natural language processing, and speech recognition areas \cite{Lecun2015Deep,He2016Deep,Graves2013Speech}. A deep neural network can learn different levels of representations on different layers of neural network \cite{Bengio2013Representation}. However, one challenge of applying deep neural networks for fraud detection is lack of sufficient labeled data.
When deep neural networks with a high dimensional input have a large number of parameters, the deep neural networks need to be trained with a large training dataset \cite{Lecun2015Deep}. Hence it is often infeasible to use the adjacency matrix of the underlying graph as inputs of deep neural network models because of the high dimension of the adjacency matrix and the small number of labeled users.

We propose a novel framework that combines spectral graph analysis with the deep neural networks.
In particular, we first project a graph to its spectral space formed by the principal eigenvectors of its adjacency matrix. The spectral space captures  the main topological information of the graph.  Each node is then mapped to a low dimensional point (called spectral coordinate) in the spectral space. We then use each node's spectral coordinate together with the aggregate information of its neighbor nodes' spectral coordinates as the input of two deep neural network models, deep autoencoder and convolutional neural network.

The advantages of our framework over past efforts are as follows. First, using both spectral graph analysis and deep neural networks, we can avoid defining graph metrics (features) to identify the difference between fraudsters and regular users. Second, the low-dimensional spectral space contains the most useful topology information of a graph. Comparing with the adjacency matrix, the dimension of spectral coordinates of nodes is much lower. Thus, using the node spectral coordinates as inputs to deep neural networks is suitable for real cases where the labeled users are limited. Moreover, most of the existing works for fraud detection focus on unsigned graphs in which there are only one type of links, while our framework covers signed networks. In order to capture both positive and negative edge information of a node in the signed graph, inputs of the two deep neural networks are composed by combining spectral coordinates of the node and its positive/negative-connected neighbors.


\section{Models}

\subsection{Framework}
\begin{figure}[h]
\centering
\includegraphics[width=.3\textwidth]{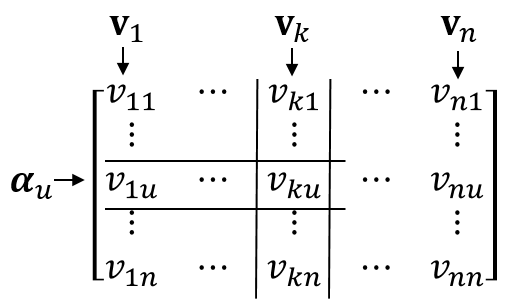}
\caption{Spectral decomposition of the adjacency matrix $\mathbf{A}$}
\label{fig:spectral}
\end{figure}

Given a signed undirected graph $G$, each node in $G$ indicates either a regular user or fraudster. The signed graph $G$ can be represented as a symmetric adjacency matrix $\mathbf{A}_{n*n}$, where $n$ is the number of nodes. In $\mathbf{A}_{n*n}$, $a_{ij}=1$ ($a_{ij}=-1$) indicates there is a positive (negative) edge between nodes $i$ and $j$ and $a_{ij}=0$ indicates no edge. $\mathbf{A}$ has $n$ real eigenvalues. Let $\lambda_i$ be the $i$-th largest eigenvalues of $\mathbf{A}$ with eigenvector $\mathbf{v}_i$, $\lambda_1 \geq \lambda_2 \geq \dots \geq \lambda_n$. The spectral decomposition of $\mathbf{A}$ is $\mathbf{A}=\sum_i {\lambda_i \mathbf{v}_i \mathbf{v}_i^T}$ (shown in Figure \ref{fig:spectral}). There usually exist $k$ leading eigenvalues that are significantly greater than the rest ones for networks. The row vector $\bm{\alpha}_u=(v_{1u},v_{2u},\dots,v_{ku})$ is the spectral coordinate of node $u$ in the $k$-dimensional subspace spanned by $(\mathbf{v}_1,\dots,\mathbf{v}_k)$.

In this work, we adopt deep autoencoder and convolutional neural network to identify fraudsters. Instead of using adjacency matrix, we adopt spectral coordinates to represent nodes as inputs to the two deep neural networks since spectral coordinates of nodes preserve the most useful structure information about nodes. Meanwhile, given a node $u$ in a graph $G$, it has neighbors in s-steps. For example, when $s=1$, the 1-step neighbors indicate the neighbors are one step away from the node $u$. Spectral coordinate of $u$'s neighbors in s-steps represent broader topological information about the node $u$. For example, spectral coordinates of 1-step neighbors have been successfully used to detect random link attacks from unsigned graphs \cite{Ying2011Spectrum}. Thus, we further adopt the spectral coordinates of node neighbors in s-steps. In a signed graph, neighbors of node $u$ can be divided into 2 categories based on their edge types, i.e., neighbors connected by positive edges and neighbors connected by negative edges. We compute the mean vector of s-step neighbors' spectral coordinates for each category, denoted as $\bm{\beta}_{u}^{s^+}$ and $\bm{\beta}_{u}^{s^-}$, where $s$ indicates the s-step neighbors. Then, given a node $u$, to capture a broad structure information of $u$, the final inputs of two deep neural networks combine the spectral coordinates of node $u$ and its two categories of neighbors in s-steps. We use a small part of labeled nodes to train the deep autoencoder and convolutional neural network. After training, the deep neural networks are able to identify fraudsters of the rest nodes in the signed graph.

\subsection{Using deep autoencoder (DAE) for fraud detection}

\begin{figure}
  \includegraphics[width=.48\textwidth]{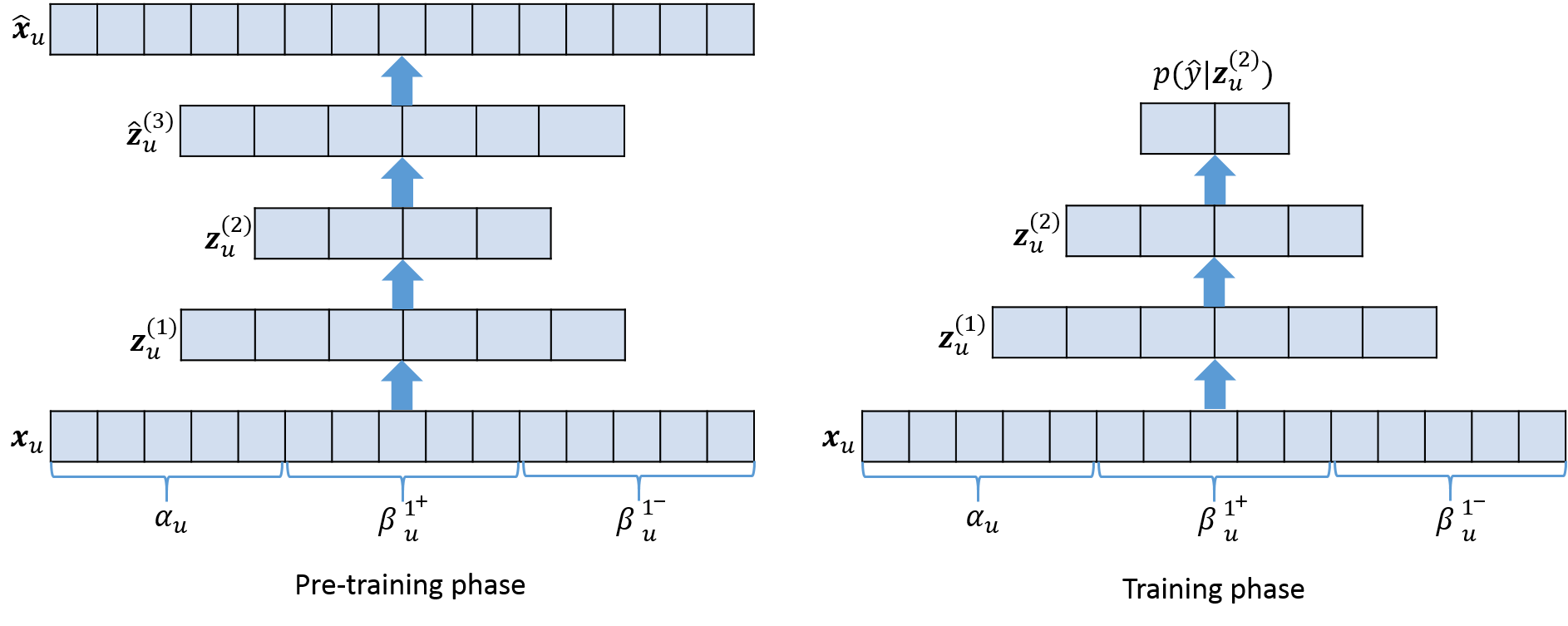}
  \caption{Architecture of DAE with spectral coordinates of node $u$ and its 1-step neighbors for fraud detection}
  \label{fig:dae}
\end{figure}

DAE stacks multiple basic autoencoder blocks hierarchically, which can capture multiple levels of representations of the input data \cite{Baldi2012Autoencoders}. We adopt spectral coordinates of nodes in a signed graph as inputs to DAE. DAE can preserve the hidden knowledge about the node from its spectral coordinate. Training DAE for fraud detection contains two phases: the pre-training phase and training phase. In the pre-training phase, DAE trains the model in an unsupervised manner. In the training phase, DAE trains the classifier and fine-tunes the whole model to predict the labels of nodes. Given the spectral coordinates of node $u$ and its two categories of neighbors in s-steps, the input $\mathbf{x}_u \in \R^{(2s+1)k}$ of DAE is defined as:
\begin{equation}
\label{eq:dae_input}
\mathbf{x}_u=[\bm{\alpha}_u ~ \bm{\beta}_{u}^{1^+} ~ \bm{\beta}_{u}^{1^-} ~ \dots ~ \bm{\beta}_{u}^{s^+} ~ \bm{\beta}_{u}^{s^-}].
\end{equation}

Given a DAE, there are $L$ encoders to compute the hidden representations $\mathbf{z}_u^{(1)}, \mathbf{z}_u^{(2)}, \dots, \mathbf{z}_u^{(L)}$ of $\mathbf{x}_u$ layer by layer. The input of the $(l+1)$-th encoder is the output of the $l$-th encoder. Specially, the input of the first encoder is $\mathbf{x}_u$. Then, there are another $L$ decoders to compute the reconstructed input $\hat{\mathbf{x}}$. The equations of encoder and decoder are shown in Equation \ref{eq:den} and \ref{eq:dde}, respectively.
\begin{equation}
\label{eq:den}
\mathbf{z}_u^{(l)} = \sigma(\mathbf{W}^{(l)} \mathbf{z}_u^{(l-1)} + \mathbf{b}^{(l)}),
\end{equation}
\begin{equation}
\label{eq:dde}
\mathbf{\hat z}_u^{(L+l)} = \sigma(\mathbf{W}^{(L+l)} \mathbf{\hat z}_u^{(L+l-1)} + \mathbf{b}^{(L+l)}),
\end{equation}
where $\mathbf{W}$, $\mathbf{b}$ are the parameters of the encoder; $\sigma$ is a nonlinear activation function.

The objective function of pre-training is to make the reconstructed input $\hat{\mathbf{x}}$ to be close to the original input $\mathbf{x}$ by minimizing the reconstruction squared error,
\begin{equation}
\label{eq:error}
\mathbi{L}(\mathbf{x}, \hat{\mathbf{x}}) = |\hat{\mathbf{x}}-\mathbf{x}|^2.
\end{equation}

After pretraining DAE, we stack $L$ encoders layer by layer to generate the hidden representation $\mathbf{z}_u^{(L)}$. $\mathbf{z}_u^{(L)}$ captures the hidden information of $\mathbf{x}_u$ since it can be used to reconstruct the input. Then, a softmax classifier is applied on top of the $\mathbf{z}_u^{(L)}$ to predict the label of node $u$.

\begin{equation}
\label{eq:softmax}
P(\hat y = c | \mathbf{z}_u^{(L)}) = \frac{\exp{(\mathbf{u}_c^T \mathbf{z}_u^{(L)} + b_c)}}{\sum_{c'=1}^C {\exp(\mathbf{u}_{c'}^T \mathbf{z}_u^{(L)} + b_{c'})}},
\end{equation}
where $C$ is number of classes, $\mathbf{u}_c$ and $b_c$ are the parameters of softmax function for the $c$-th class. The parameters of softmax and deep autoencoder are trained and fine-tuned by minimizing the cross entropy loss function,
\begin{equation}
\label{eq:loss}
\mathbi{L} = -\frac{1}{N} \sum_{i=1}^{N} y_i * \log(P(\hat{y}_i)),
\end{equation}
where $y_i$ is the true class of the $i$-th input, and $N$ is the number of training data. Figure \ref{fig:dae} shows the architecture of DAE with spectral coordinate of node $u$ and its 1-step neighbors.

\subsection{Using convolutional neural network (CNN) for fraud detection}

\begin{figure}
  \includegraphics[width=.45\textwidth]{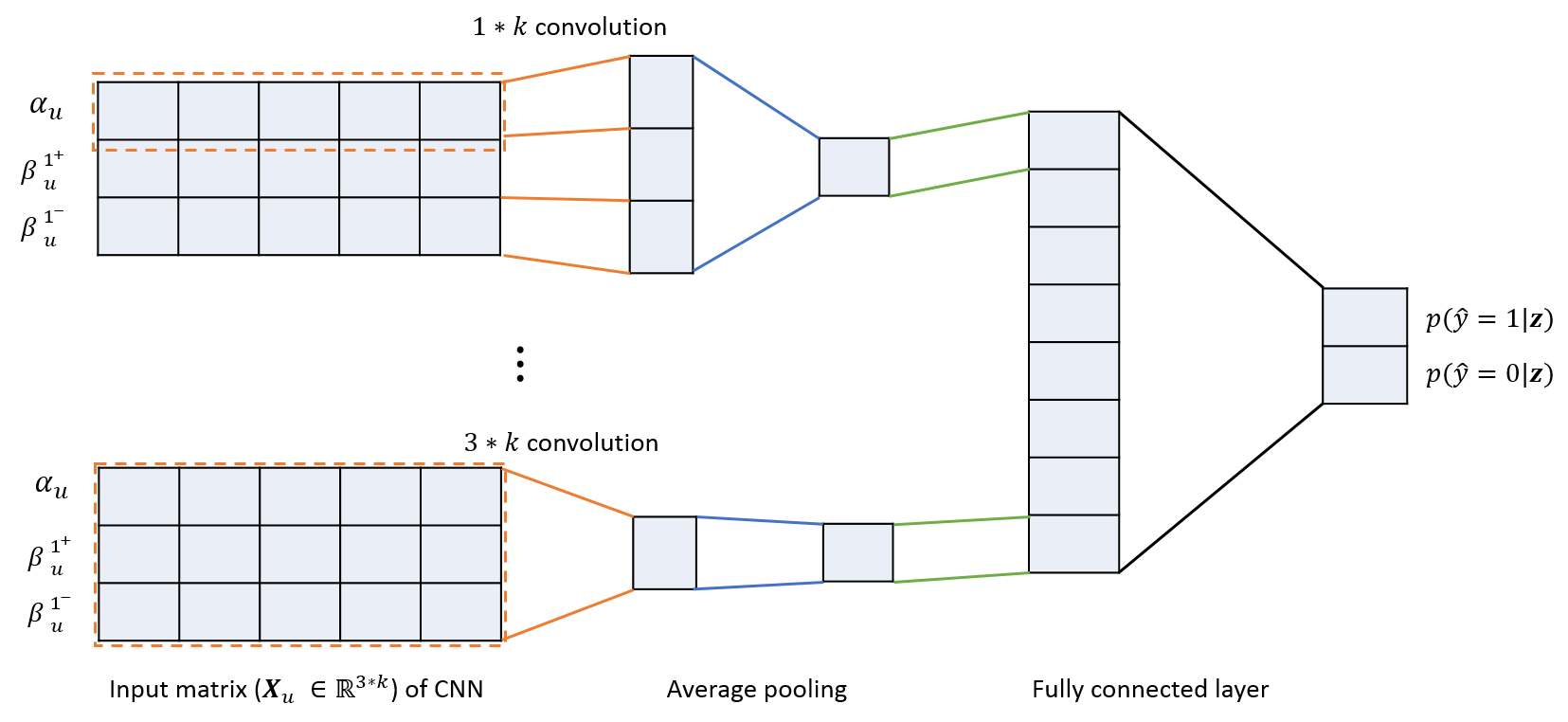}
  \caption{Architecture of CNN with spectral coordinates of node $u$ and its 1-step neighbors for fraud detection}
  \label{fig:cnn}
\end{figure}
Convolutional neural networks are widely-used in computer vision area\cite{Krizhevsky2012Imagenet}. A basic convolutional neural network is composed by a convolution operation and a pooling operation.
Given a node $u$, the input of CNN $\mathbf{X}_u \in \R^{(2s+1)*k}$ is represented as
\begin{equation}
\label{eq:cnn_input}
    \mathbf{X}_u = [\bm{\alpha}_u; \bm{\beta}_{u}^{1^+}; \bm{\beta}_{u}^{1^-}; \dots; \bm{\beta}_{u}^{s^+}; \bm{\beta}_{u}^{s^-}],
\end{equation}
where ; indicates the vertical concatenation of two vectors.
A convolution operation which involves a filter $\mathbf{W} \in \R^{m*k}$ is applied on a sub-matrix of $\mathbf{X}_u$ with $m$ continuous rows to generate a hidden feature:
\begin{equation}
\label{eq:conv}
  h_j = \sigma((\mathbf{W}*\mathbf{X}_u)_{j:j+m-1}+b),
\end{equation}
where $\sigma$ indicates a nonlinear function; $*$ is a convolution operation; $b$ is a bias parameter; and $m \leq 2s+1$. Initially, the filter is partially connected to the input matrix $\mathbf{X}_u$. Then, the filter slides through the whole input matrix $\mathbf{X}_u$ and generates a feature vector $\mathbf{h}=[h_1,\cdots,h_{2s-m+2}]$. After that, an average pooling operation, which calculates the average value of the feature vector $\mathbf{h}$, is used to capture all discriminative features of $\mathbf{X}_u$. The average pooling operation is defined as $z=mean(\mathbf{h})$.

For example, when the filter size $m$ is 1, the convolution operation generates a feature vector $\mathbf{h} \in \R^{2s+1}$ which contains the hidden information of node $u$ and its all positive/negative-connected neighbors in s-steps. Then, an average pooling operation is applied on $\mathbf{h}$ to get the hidden feature $z$ by calculating the average value of $\mathbf{h}$ as the output of current filter. When the filter size $m>1$, the model follows the similar procedure to generate hidden features from $\mathbf{X}_u$.

Meanwhile, because one hidden feature vector $\mathbf{h}$ is generated by one filter, the feature vector detects the same feature from different locations of $\mathbf{X}_u$. To identify different aspects of features from $\mathbf{X}_u$, the model applies multiple filters with different sizes of $m$ to generate different feature vectors. After applying the pooling operation on each feature vector, the model encodes the input $\mathbf{X}_u$ to a representation vector $\mathbf{z} = [z_1, z_2, \cdots , z_q]$ , where $q$ is the number of feature vectors generated by $q$ different filters. For each filter size $m$, CNN generates the same number of feature vectors. We then apply a softmax classifier on $\mathbf{z}$ to predict the node label $P(\hat y = k |\mathbf{z})$ by Equation \ref{eq:softmax}. The whole model is trained by minimizing the loss function shown in Equation \ref{eq:loss}. The architecture of CNN with spectral coordinates of node $u$ and its 1-step neighbors is shown in Figure \ref{fig:cnn}.

\section{Experiments}
To evaluate the effectiveness and efficiency of our approach, we conduct experiments on a signed graph for fraud detection.

\subsection{Experimental Setup}

{\noindent \bf Datasets.}
We conduct our evaluation on a signed network, \textit{WikiEditor}, which is extracted from the UMD Wikipedia dataset\cite{Kumar2015Vews}. The dataset is composed by 17015 vandals and 17015 benign users who edited the Wikipedia pages from Jan 2013 to July 2014. Different from benign users, vandals edit articles in a deliberate attempt to damage Wikipedia. One edit may be reverted by bots or editors. Hence, each edit can belong to either revert or no-revert category. WikiEditor is built based on the co-edit relations. In particular, a positive (negative) edge between users $i$ and $j$ is added if the majority of their co-edits are from the same category (different categories). We remove those edits on meta pages (i.e., with titles containing ``User:'', ``Talk:'', ``User talk:'', ``Wikipedia:'') because the editings on those pages are not reverted by bot or administrators. We further remove from our signed network those users who do not have any co-edit relations with others. In WikiEditor, each user is clearly labeled as either benign or vandal. Hence, we can evaluate our models for fraud detection on WikiEditor.

\begin{table}[h]
\centering
\caption{Statistics of WikiEditor}
\label{tb:wiki}
\begin{adjustbox}{max width=0.48\textwidth}
\begin{tabular}{|c|c|c|}
\hline
           & \# of Users (+,-)   & \# of Links (+,-)      \\ \hline
WikiEditor & 18992 (6086, 12906) & 81316 (52139, 29177) \\ \hline
\end{tabular}
\end{adjustbox}
\end{table}

{\noindent \bf Experimental settings.}
In our experiments, after projecting the signed graph to the spectral space, we first normalize the spectral coordinates of nodes. We then combine nodes and their 1-step neighbors' spectral coordinates as inputs to deep neural networks. DAE contains two encoders. The dimensions of two encoders are 128 and 64. In CNN, the filter size $m \in [1,2,3]$ and the number of filters $q$ is 300. The training epochs of DAE and CNN models are 30 with early stopping. We randomly sample different percentages of nodes for training and use the rest of the nodes for testing. Since the labeled fraudsters in real cases are usually small, we only adopt a small percentage of nodes as training data. We use the accuracy to evaluate the performance of different approaches for vandal detection. We report the mean values of 10 different runs by sampling different training data.

{\noindent \bf Baselines.}
We compare deep neural networks with two classical classifiers, k-NN and SVM. In k-NN, we set $k=3$. In SVM, we adopt the RBF kernel and set the regularization parameter $c=1$. The inputs of k-NN and SVM are the same as DAE.

\begin{table}[htbp!]
\centering
\caption{Accuracy of vandal detection with various sizes of training dataset}
\label{tb:accuracy}

\begin{adjustbox}{max width=0.5\textwidth}

\begin{tabular}{|c|c|c|c|c|c|}
\hline
\multirow{2}{*}{Input}           & \multirow{2}{*}{Algorithm}     & \multicolumn{4}{c|}{Ratio of the training dataset}                               \\ \cline{3-6}
                                 &           & 5\%                & 10\%               & 15\%               & 20\%               \\ \hline
\multirow{4}{*}{$\mathbf{A}$}    & k-NN      & 66.16\%            & 68.82\%            & 69.66\%            & 74.00\%           \\ \cline{2-6}
                                 & SVM       & 67.81\%            & 67.82\%            & 67.88\%            & 67.92\%            \\ \cline{2-6}
                                 & DAE       & 76.31\%            & 78.55\%            & 79.56\%            & 80.59\%                  \\ \cline{2-6}
                                 & CNN       & \textbf{76.70\%}   & \textbf{78.95\%}   & \textbf{80.09\%}   & \textbf{81.33\%}  \\ \hline  \hline
\multirow{4}{*}{$\mathbf{x}_u$}  & k-NN      & 76.60\%            & 77.38\%            & 77.83\%            & 78.19\%            \\ \cline{2-6}
                                 & SVM       & 71.60\%            & 80.40\%            & 80.82\%            & 81.15\%            \\ \cline{2-6}
                                 & DAE       & \textbf{80.89\%}   & 81.13\%   		   & 81.45\%            & 81.92\%           \\ \cline{2-6}
                                 & CNN       & 80.57\%            & \textbf{81.40\%}   & \textbf{82.02\%}   & \textbf{82.61\%}  \\  \hline
\end{tabular}
\end{adjustbox}
\end{table}

\subsection{Results}
We first compare each model with different sizes of training datasets using spectral coordinates and the adjacency matrix. The default dimension of spectral coordinate $k$ is 30. When using the adjacency matrix $\mathbf{A}$, the input of each algorithm is each row of $\mathbf{A}$. In this scenario, CNN has only one size filter $\mathbf{w} \in ^{1*n}$.

{\noindent \bf Deep neural networks vs. Baselines.}
Table \ref{tb:accuracy} shows the accuracy of vandal detection using the adjacency matrix $\mathbf{A}$ and spectral coordinates $\mathbf{x}_u$. We can observe that DAE and CNN models outperform the baselines significantly in all settings at the $10^{-8}$ level with a t-test.
In the spectral space, the performance of SVM has a big jump when the percentage of training data increasing from 5\% to 10\%. In the adjacency matrix space, the performances of SVM do not improve while increasing the size of training data. It indicates that the SVM cannot be well-trained with a small training dataset, especially when the input has high dimensions. The accuracies of k-NN increase steadily while increasing the training data, but the accuracies of k-NN are also much worse than our DAE and CNN.

{\noindent \bf Spectrum vs. Adjacency matrix.}
In Table \ref{tb:accuracy}, we further observe that using the spectral coordinates as inputs ($\mathbf{x}_u$) achieves significantly better performance than using the adjacency matrix, especially when the percentages of training data are 5\% and 10\%.
Meanwhile, in the spectral space, when the percentage of training dataset is 5\%, DAE performs better than CNN since DAE pre-trains the model first. The pre-training phase of DAE encodes the information of nodes into hidden layers, which make the classifier  predict the node labels with small training data. On the contrary, when the percentage of training dataset is larger than 5\%, CNN outperforms DAE in the spectral space. It indicates CNN has better performance with enough training data. In the adjacency matrix space, the performances of DAE are worse than CNN with various sizes of training data. This is because DAE has much more parameters than CNN when using the adjacency matrix. DAE cannot be well-trained in a high-dimensional space with a small training dataset.

{\noindent \bf Effect of the dimension of spectral coordinate $\bm{k}$.}
Table \ref{tb:sensitivity} compares the deep neural networks with k-NN and SVM on varying the dimension of spectral coordinate $k$. In this experiment, we use 20\% of nodes as training data and the rest of nodes as testing data. When inputs of algorithms are $\mathbf{x}_u$, we can observe that DAE and CNN models outperform classical classifiers with various dimensions of spectral coordinates. We can further discover that DAE and CNN also achieve the most stable accuracy with various dimensions of spectral coordinates. However, the performance of SVM significantly drops while increasing the dimension of the spectral coordinate. This is because both DAE and CNN learn the hidden representations of nodes from their spectral coordinates. The hidden representations of nodes are useful for predicting the labels.

{\noindent \bf Neighbor inclusion vs. Neighbor exclusion.}
In our experiment, the inputs of DAE and CNN combine spectral coordinates of nodes and their 1-step neighbors. We further compare the performance of algorithms that adopt the node spectral coordinate with and without combining the 1-step neighbors' spectral coordinates as inputs in Table \ref{tb:sensitivity}. We can observe that when using the information of the 1-step neighbors' spectral coordinates, the accuracies of all algorithms achieve around 1\%-2\% improvement. Therefore, combining the information of node neighbors can improve the performance of fraud detection.

{\noindent \bf Execution time.}
We also compare the execution time of deep neural networks using spectral coordinates and the adjacency matrix.
The DAE and CNN models are trained on a Nvidia Tesla K20 GPU. We observe that when the ratio of training data is 20\%, the training time of each epoch for CNN and DAE models with adjacency matrix is 2 seconds. On the contrary, the training time of each epoch for CNN and DAE models with spectral coordinates is less than 1 second. Therefore, using the spectral coordinates with low dimension is also more efficient than using the adjacency matrix.

\begin{table}[]
\centering
\caption{The accuracy of vandal detection with various dimensions of spectral coordinate $k$ when 20\% of nodes are used as the training dataset. We further compare algorithms using node spectral coordinate with/without combining neighbor spectral coordinates as inputs to algorithms.}
\label{tb:sensitivity}
\begin{adjustbox}{max width=0.48\textwidth}
\begin{tabular}{|c|c|c|c|c|c|c|}
\hline
\multirow{2}{*}{Input}           & \multirow{2}{*}{Algorithm} & \multicolumn{5}{c|}{Dimension of spectral coordinate $k$}                              \\ \cline{3-7}
                                 &           & 10               & 20               & 30               & 40               & 50 \\ 				\hline
\multirow{4}{*}{$\mathbf{x}_u$}  & k-NN      & 79.90\%          & 78.85\%          & 78.19\%          & 78.33\%          & 77.52\%          \\ \cline{2-7}
                                 & SVM       & 81.00\%          & 81.40\%          & 81.15\%          & 80.70\%          & 76.28\%          \\ \cline{2-7}
                                 & DAE       & 81.86\%          & 81.65\%          & 81.92\%          & 81.87\%          & 81.61\%          \\ \cline{2-7}
                                 & CNN       & \textbf{82.24\%} & \textbf{82.26\%} & \textbf{82.61\%} & \textbf{82.42\%} & \textbf{82.47\%} \\ \hline \hline
\multirow{4}{*}{$\bm{\alpha}_u$} & k-NN      & 78.00\%          & 77.49\%          & 76.75\%          & 76.55\%          & 76.92\%          \\ \cline{2-7}
                                 & SVM       & 79.79\%          & 79.65\%          & 80.19\%          & 80.31\%          & 77.94\%          \\ \cline{2-7}
                                 & DAE       & 80.47\%          & 81.10\%          & 81.17\%          & 81.20\%          & \textbf{81.40\%} \\ \cline{2-7}
                                 & CNN       & \textbf{80.52\%} & \textbf{81.22\%} & \textbf{81.48\%} & \textbf{81.44\%} & \textbf{81.39\%} \\ \hline
\end{tabular}
\end{adjustbox}
\end{table}

\section{Conclusions}
We have presented a novel framework that applies deep neural networks on the spectral space of a signed graph to identify frauds. In particular, we first conduct graph spectral projections on a signed graph to obtain node spectral coordinates. The node and its s-step neighbors' spectral coordinates are combined together as inputs to the deep autoencoder and convolutional neural network models for fraud detection. The experiment results show that both deep neural networks achieve promising results on fraud detection. Our empirical evaluation further shows that combining the information of node neighbors can improve the effectiveness of deep neural networks on fraud detection.

\begin{acks}
The authors acknowledge the support from the 973 Program of China (2014CB340404) and China Scholarship Council to Shuhan Yuan and from National Science Foundation  to Xintao Wu (1564250), Jun Li (1564348) and Aidong Lu (1564039). This research was conducted while Shuhan Yuan visited University of Arkansas.
\end{acks}

%


\end{document}